\documentclass[reprint,aps,prl,superscriptaddress,showkeys,amsmath,amssymb,letterpaper]{revtex4-2}

\def\wn/{\,cm$^{-1}$}
\def\area/{\,cm$^{-2}$}

\newcommand{\fts}[1]{FeTe$_{1-#1}$Se$_{#1}$}

\def\edc/{\ensuremath{\mathbf{E}}}
\def\eac/{\ensuremath{\mathbf{E}_\omega}}
\def\bdc/{\ensuremath{\mathbf{B}}}
\def\hdc/{\ensuremath{\mathbf{H}}}
\def\bac/{\ensuremath{\mathbf{B}_\omega}}
\def\hac/{\ensuremath{\mathbf{H}_\omega}}

\newcommand{\vect}[1]{\ensuremath{\mathbf{#1}}}
\newcommand{\mat}[1]{\ensuremath{\hat{\mathbf{#1}}}}

\usepackage{graphicx}   
\usepackage{bm}                 
\usepackage{amsmath}
\usepackage{amsfonts}
\usepackage{revsymb}
\usepackage{color}
\usepackage{soul}
\usepackage[colorlinks=true,urlcolor=blue,citecolor=blue,linkcolor=blue,bookmarks=false,pdfstartview={FitH}]{hyperref}

\usepackage{relsize}

\graphicspath{{pdf/}}

\begin{document}

\title{Chirality-Resolved Spectroscopy of Caroli–de Gennes–Matricon sSates in Multiband \fts{x} Superconductors}
\author{T.\,R\~o\~om}
\email{toomas.room@kbfi.ee}
\affiliation{National Institute of Chemical Physics and Biophysics, Akadeemia tee 23, 12618 Tallinn, Estonia}
\author{A.\,Glezer\,Moshe}
\affiliation{National Institute of Chemical Physics and Biophysics, Akadeemia tee 23, 12618 Tallinn, Estonia}
\author{R.\,Nagarajan}
\affiliation{National Institute of Chemical Physics and Biophysics, Akadeemia tee 23, 12618 Tallinn, Estonia}
\author{U.\,Nagel}
\affiliation{National Institute of Chemical Physics and Biophysics, Akadeemia tee 23, 12618 Tallinn, Estonia}
\author{Hee\,Taek\,Yi}
\affiliation{Department of Physics and Astronomy, Rutgers University, Piscataway, New Jersey 08854, USA}
\author{Seongshik\,Oh}
\affiliation{Department of Physics and Astronomy, Rutgers University, Piscataway, New Jersey 08854, USA}
\author{G.\,Blumberg}
\email{girsh@physics.rutgers.edu}
\affiliation{National Institute of Chemical Physics and Biophysics, Akadeemia tee 23, 12618 Tallinn, Estonia}
\affiliation{Department of Physics and Astronomy, Rutgers University, Piscataway, New Jersey 08854, USA}

\date{January 6, 2026}

\begin{abstract}
We employ terahertz Faraday magneto-optical spectroscopy to probe the relaxation dynamics of quantized helical Caroli--de Gennes--Matricon (CdGM) states in epitaxial FeTe$_{1-x}$Se$_x$ thin films, nodeless multiband superconductors with short coherence lengths in the moderately clean limit.
By exploiting polarization-selective optical transitions, we directly resolve the helicity and band origin of vortex-core quasiparticles.
We observe resonant transitions between long-lived CdGM states induced by radiation with opposite helicity for electronlike and holelike bands.
This enables independent, band-resolved determination of quasiparticle lifetimes, vortex masses, coherence lengths, and upper critical fields, and reveals their systematic evolution with isovalent substitution $x$.
The results establish terahertz magneto-optics as a direct probe of helical vortex-core excitations and provide dynamical evidence for multiband CdGM states in iron-based superconductors.
\end{abstract}


\maketitle

\emph{Introduction.} 
In the Abrikosov phase of type-II superconductors, vortex cores control low-energy dissipation and electrodynamics. 
When the quasiparticle spectrum of the bulk superconductor is fully gapped, the only in-gap excitations reside within vortex cores. 
In the classic picture of Bardeen and Stephen, the core is treated as a normal-metal region whose carriers respond to electromagnetic fields as in the normal state~\cite{Bardeen-Stephen}. 
This description applies to superconductors in the {\it dirty} limit, where the electronic mean free path $l$ is much shorter than the superconducting coherence length $\xi_0$, characteristic size of the vortex core.

\begin{figure}[t]
     \includegraphics[width=0.95\columnwidth]{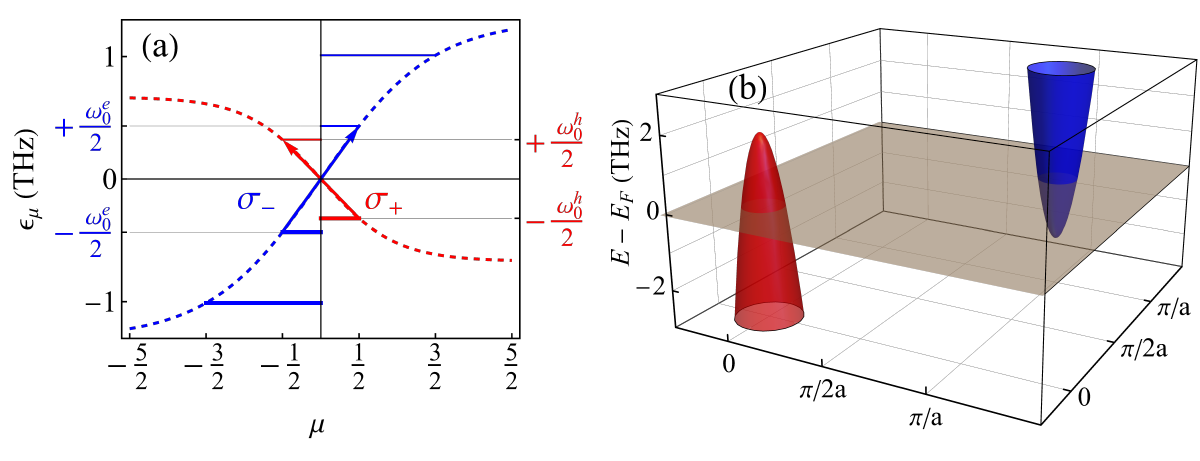}
     \vspace{-2mm}
    \caption{\label{fig:BZ_and_mu}
     (a) Energies of the CdGM states for electrons (blue) and holes (red), $\epsilon_\mu^k = \Delta_k \tanh(\mu \Delta_k/ E_F^k)$~\cite{Bardeen-Tewordt}.
    $\Delta_e=1.3$\,THz, $\Delta_h=0.67$\,THz \cite{Homes2015} and $E_F^k=\Delta_k^2/2\pi\omega_0^k$ for sample $x=0.38$, Table\,\ref{tab:fit_values}.
       Arrows show the transitions from the highest occupied to the lowest unoccupied state, blue for left-handed  and red for right-handed radiation. 
       (b) A cartoon of two-band structure for iron-based superconductors with an electron band (blue) at the Brillouin zone $M$ point ($m_e^*=3.4m_0$, $E_F=1.7$\,THz; $m_0$ is the bare electron mass) and hole band (red) at the $\Gamma$ point ($m_h^*=3.3m_0$, $E_F=-1.9$\,THz). 
     \vspace{-5mm}
          }
\end{figure}

In {\it clean} $s$-wave superconductors, where $l>\xi_0$, vortex cores host discrete bound states first described by Caroli, de Gennes, and Matricon (CdGM)~\cite{Caroli1964}. 
Their energies,
$\epsilon_\mu \approx \mu \Delta^2/E_F \approx 2\mu \hbar^2/(m^*\xi_0^2)$,
are indexed by half-odd-integer angular momentum $\mu$~\cite{Bardeen-Tewordt,Janko-Shore}, Fig.~\ref{fig:BZ_and_mu}(a). 
Here, $\Delta$ is the superconducting gap, $E_F$ is  Fermi energy and   $m^*$ is quasiparticle band mass. 
At low temperatures, negative-energy states are occupied while positive-energy states remain empty~\cite{chem_potential}. 
Transitions between CdGM levels couple to the condensate via Andreev scattering and, together with impurity scattering, determine the dynamical response and dissipation of the vortex core excitations~\cite{Bardeen-Tewordt}.

In conventional superconductors, the long coherence length and small ratio $\Delta/E_F$ render the CdGM level spacing smaller than thermal and lifetime broadening, precluding spectroscopic resolution of individual levels. 
Accessing the {\it quantum} regime therefore requires superconductors with short coherence lengths. 
Although cuprate high-$T_{c}$ superconductors satisfy this condition, their nodal order parameter produces a quasicontinuous vortex-core spectrum that obscures discrete bound states~\cite{Renner2017,Karrai1992,Hsu1993,Hsu1994,Choi1994,Choi1995,Kopnin1995,Volovik_1997,Lihn1997,Nikolic2006,Tesar2021,Tesar2024}. 
Iron-based superconductors provide a favorable alternative. 
In particular, FeTe$_{1-x}$Se$_x$ combines large $\Delta/E_F$ ratio~\cite{Rinott2017} with nodeless $s^{\pm}$ bulk order parameter that changes sign between Fermi-surface pockets but remains fully gapped on each~\cite{Mazin2010}. 
In a minimal description, the Fermi surface consists of a hole pocket at the Brillouin-zone center and an electron pocket at the zone corner, Fig.\,\ref{fig:BZ_and_mu}(b). 
We will restrict discussion to the two lightest bands for which the CdGM states are at higher frequency~\cite{fermi}.

Multiband superconductivity introduces additional richness to vortex physics~\cite{Suhl1959,Zehetmayer2013,Lin2014}. 
Distinct gaps on different Fermi-surface sheets lead to multiple characteristic core sizes, while phase frustration between condensates can support fractional vortices and nonmonotonic intervortex interactions~\cite{Babaev2002,Babaev2005,Babaev2011,Babaev2025}. 

Recent scanning tunneling spectroscopy (STS) experiments have resolved individual CdGM states in iron-based superconductors~\cite{HDing2018,Hanaguri2019,Chen2018}. 
While STS resolves the static local density of states in vortex cores~\cite{HHess1990,Renner2017,HDing2018,Hanaguri2019,Chen2018,Yan2026}, it is insensitive to the handedness imposed by magnetic field and therefore cannot access the vortex-core quasiparticle helicity.
In contrast, circularly polarized terahertz photons obey strict angular-momentum selection rules, enabling direct, dynamical probes of CdGM helicity. 

Because photons carry angular momentum $\hbar$, optical transitions between CdGM states obey strict helicity selection rules.
For electron-like carriers, left-handed circularly polarized radiation induces resonant transitions between the occupied $\mu=-\tfrac{1}{2}$ and unoccupied $\mu=+\tfrac{1}{2}$ states at frequency $\omega_0=(\epsilon_{1/2}-\epsilon_{-1/2})/\hbar$, Fig.\,\ref{fig:BZ_and_mu}(a).
By charge-conjugation and time-reversal symmetry, hole-like carriers absorb right-handed polarization, corresponding to transitions between the $\mu=+\tfrac{1}{2}$ and $\mu=-\tfrac{1}{2}$ states \cite{Janko-Shore}.
This polarization selectivity of absorption directly links the Faraday response to the band character and helicity of vortex-core quasiparticles.
The CdGM transition frequency $\omega_0$ is of the same order as the cyclotron frequency $\omega_{c} = e B_{c2}/m^*$ of elementary charge $e$ in  upper critical field $B_{c2}$~\cite{Kohn1961,Kopnin1995,Kopnin2001}.

\begin{figure}[b]
     \vspace{-3mm}
   \includegraphics[width=0.95\columnwidth]{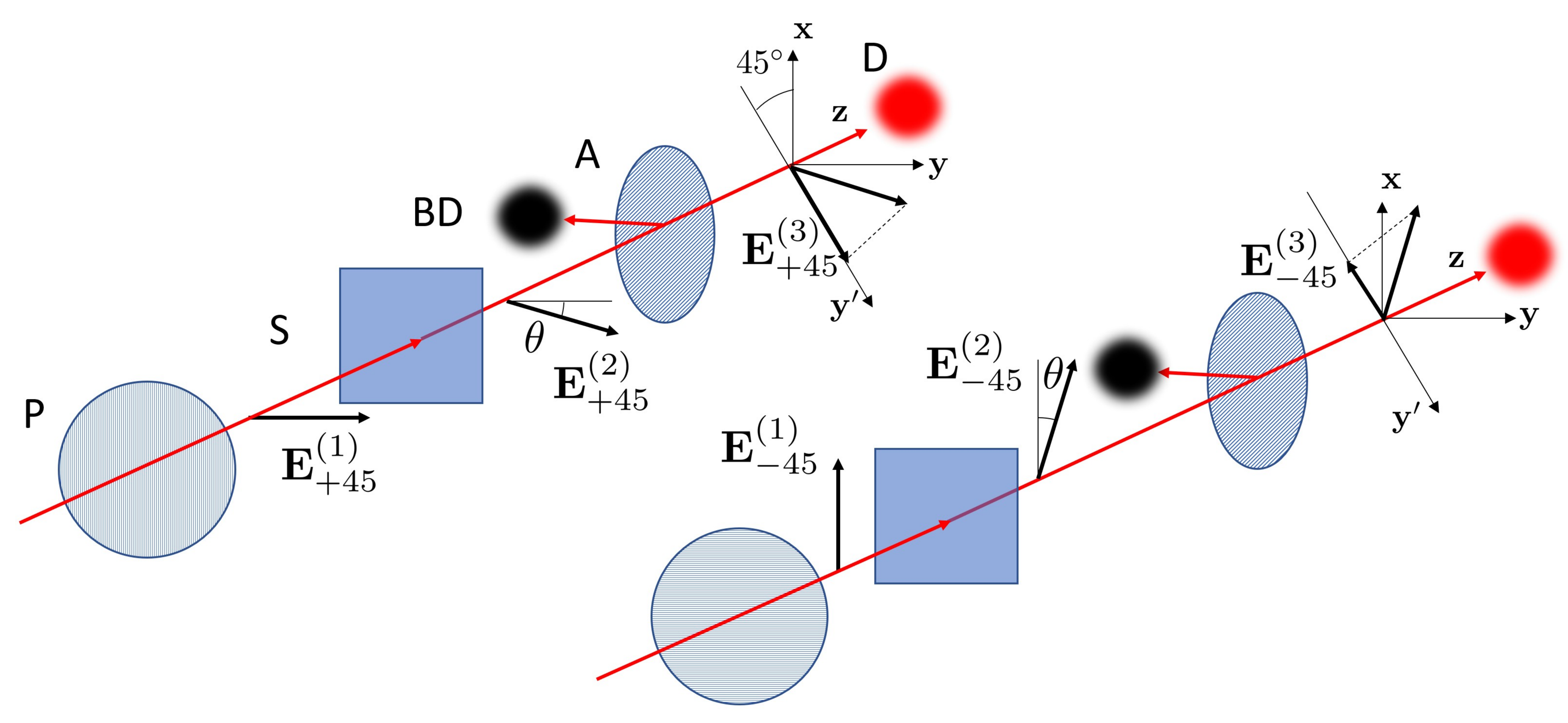}
    \vspace{-2mm}
    \caption{\label{fig:Faraday_pm45}
        The schematics of Faraday rotation measurement setup showing  rotation $\theta$. 
        P - polarizer, S - sample, A - analyzer tilted by $45^\circ$ from the beam direction to reflect the wrong polarization into beam dump BD, D - detector.
        Left: incident electric field in $+45^\circ$ orientation, $\vect{E}_{+45}^{(1)}\parallel \vect{y}$; 
        Right: in $-45^\circ$ orientation, $\vect{E}_{-45}^{(1)}\parallel \vect{x}$.
        $\vect{E}_{\pm45}^{(2)}$ and $\vect{E}_{\pm45}^{(3)}$ are electric field  transmitted through the sample and after passing the analyzer with its polarization axis along $\vect{y}'$, respectively.   
        For a small clockwise  $\theta$ the amplitudes at the detector are $\vect{E}_{+45}^{(3)}>\vect{E}_{-45}^{(3)}$.}
\end{figure}
The asymmetry in absorption leads to distinct and experimentally identifiable refractive indices for the light with two circularly polarized (helical) basis vectors, ${e}_{\pm} = ({e}_x \pm i{e}_y)/\sqrt{2}$, corresponding to light propagating along the \vect{z} axis, parallel to the externally applied magnetic field $\vect{B}$: 
\begin{equation}\label{eq:N_index}
N^{\pm}(\omega, B) = \sqrt{\epsilon_{\infty} + i  \sigma^{\mp}(\omega, B)/\epsilon_0\omega }, 
\end{equation}
where $\sigma^{\pm}(\omega,B)$ is the optical conductivity for the corresponding polarization, $\epsilon_0$ the permittivity of the vacuum, and $\epsilon_{\infty}$ the high-frequency dielectric constant.
This difference can be measured by the Faraday rotation: 
when linearly polarized light travels inside a superconductor over distance $d$, it acquires ellipticity with the major axis of the ellipse rotated by the angle $\theta_F$ with respect to the incident light polarization~\cite{Argyres1955}:  
\begin{equation}\label{eq:f_rot}
\theta_\text{F}(\omega,B) = (\omega d/2c)\,\Re\,(N^{+}-N^{-}), 
\end{equation}
where $c$ is the speed of light. 
Thus, the Faraday rotation and ellipticity provide direct access to the difference in optical response of helical vortex-core excitations.

In this Letter, we employ terahertz Faraday magneto-optical spectroscopy to probe vortex-core dynamics in thin films of FeTe$_{1-x}$Se$_x$, a moderately clean limit nodeless multiband superconductor with short coherence length.
The measurements access both the magnitude and the helicity of the vortex response, allowing us to resolve electron- and hole-band contributions to the CdGM spectrum and study their evolution with composition $x$.  
Furthermore, the resonant transitions between CdGM states enable independent estimates of the upper critical field and coherence length for each of the condensates. 
We also find that the effective Verdet parameter for type II superconductors is significant. 
These results establish THz magneto-optics as a probe of vortex-core dynamics and provide direct evidence of helical multiband CdGM states in FeTe$_{1-x}$Se$_x$.

\emph{Magneto-optics in multi-band \fts{x} superconductor.} 
The schematics of the Faraday rotation measurement setup is shown in Fig.\,\ref{fig:Faraday_pm45}.
Radiation from the Hg-arc lamp passing through the Martin-Puplett interferometer is incident on the sample, which is placed in the magnetic field of a superconducting magnet.
The polarizer before the sample was set at $\phi=\pm45^\circ$ relative to the analyzer after the sample. 
Transmitted intensity spectra, $S_{\pm45^\circ}(\omega,\pm B)\propto |\vect{E}_{\pm45}^{(3)}(\omega,\pm B)|^2$, were acquired for both field directions, $\vect{B}=\pm B \vect{z}$, using a 0.3\,K bolometer.
The Faraday rotation angle was calculated by
\begin{equation}\label{eq:full_protocol}
    \theta_\text{F}(\omega,B) = \frac{1}{4}\,\frac{r_{+45}-r_{-45}}{r_{+45}+r_{-45}}, 
\end{equation}
where $r_{\pm45}=S(\pm45^\circ,+B)/S(\pm45^\circ,-B) $~\cite{Levallois2015,SM}.

\begin{figure}[t]
    \includegraphics[width=.95\columnwidth]{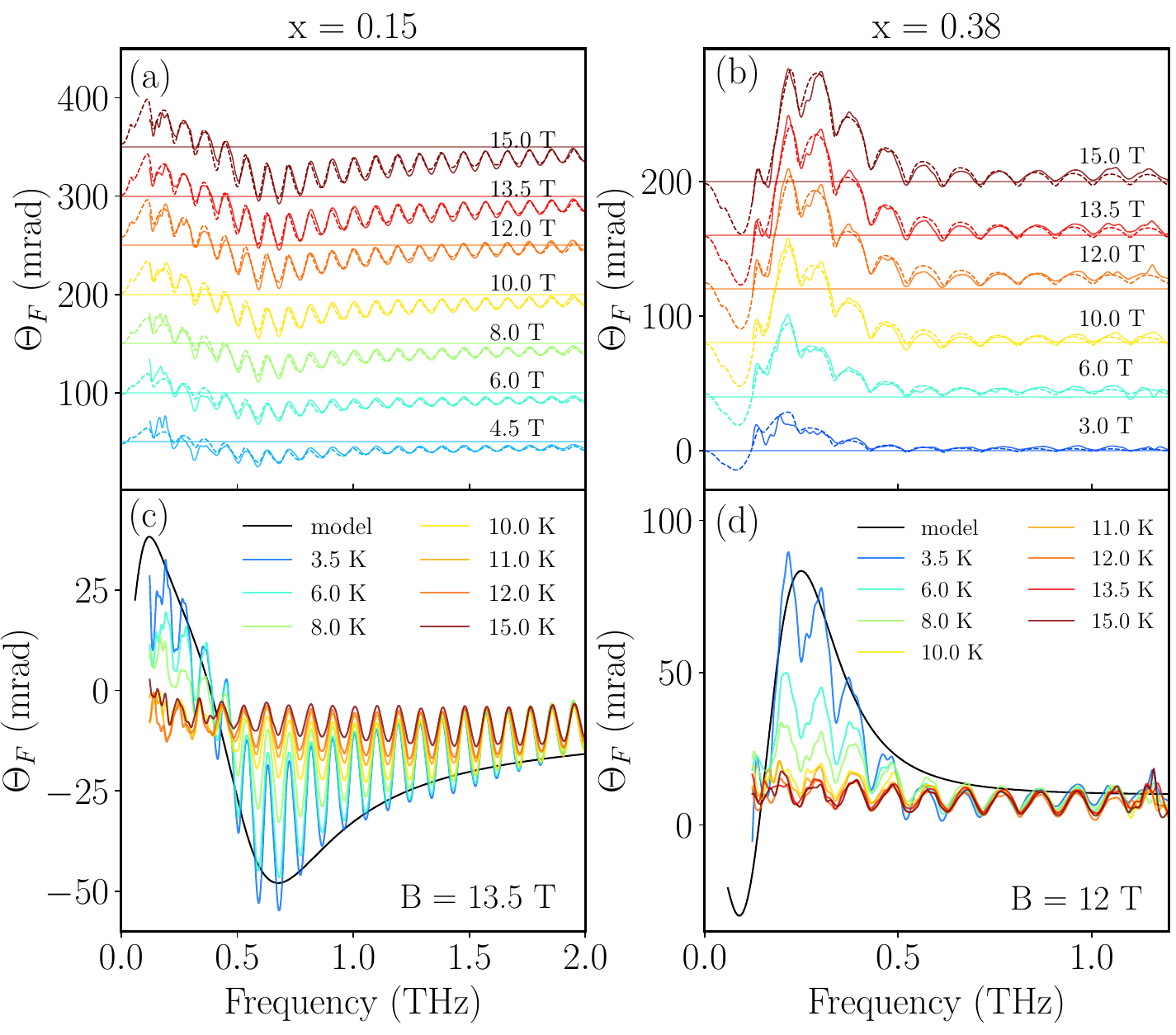}
    \caption{\label{fig:thetaF_B_T_dep}
       Magnetic field and temperature dependence of Faraday rotation spectra. 
       (a) $x=0.15$ and (b) $x=0.38$: magnetic field dependence at 3.5\,K. Solid lines are experimental data,  dashed lines are fits. 
       The curves are offset by 40\,mrad and horizontal line marks zero rotation for each curve.  
       (c) $x=0.15$ and (d) $x=0.38$: temperature dependence measured at 13.5 and 12\,T correspondingly. 
       The black lines in (c) and (d) show $\theta_F$ of a freestanding 40\,nm thick \fts{x} film at 3.5\,K, Eq.\,(\ref{eq:fit_function}), calculated using transmission $t_\pm$, Eq.\,(\ref{eq:slab}), with parameters from Table\,\ref{tab:fit_values}. 
       } 
        \vspace{-5mm}
\end{figure}

In Fig.\,\ref{fig:thetaF_B_T_dep} we show the magnetic-field and temperature  dependence of the Faraday rotation for two 40\,nm thick epitaxially grown \fts{x} superconducting films with $T_{c} \approx 13$\,K~\cite{SOh_2021,SOh_2025,TSS}.
The interference fringes in the spectra are due to Fabry-Perot resonances within the $\mathrm{Al_2O_3}$ substrate (see End Matter).

The rotation spectra for sample $x=0.15$ are negative above 0.3\,THz and positive below, Fig.\,\ref{fig:thetaF_B_T_dep}(a). 
On the contrary, the rotation for sample $x=0.38$ is mostly positive, except below 0.2\,THz, Fig.\,\ref{fig:thetaF_B_T_dep}(b).
This sign change in rotation as a function of frequency implies that both electron and hole CdGM states contribute. 

The rotation for sample $x=0.38$ reaches a maximum at about 0.3\,THz where  the rotation is close to  75\,mrad in 15\,T field, Fig.\,\ref{fig:thetaF_B_T_dep}(b)~\cite{max_rot}. 
The corresponding value for effective Verdet parameter reaches $10^5$\,rad\,(T\,m)$^{-1}$~\cite{Verdet_par}. 
This enhancement arises from the concentration of magnetic field and from the electric-dipole coupling driven optical response within vortex cores hosting quantized CdGM excitations.

The rotation is suppressed upon approaching $T_c$, Fig.\,\ref{fig:thetaF_B_T_dep}(c,d), because the expanding coherence length $\xi(T)$ decreases the CdGM resonance frequency.
The rotation above $T_c$ is negative for the $x=0.15$ and positive for the $x=0.38$ sample, consistent with the sign of the Hall coefficient $R_H$, showing that for the $x=0.15$ sample the normal state transport is dominated by holes, while for the $x=0.38$ sample it is dominated by electrons, see Fig.\,\ref{fig:Rxx_RH} in End Matter. 

\emph{Dynamic vortex mass.} 
To describe vortex dynamics in the Abrikosov phase, Kopnin {\it et. al.} introduced a frequency-dependent effective vortex-mass tensor per unit length, $\mat{M}$, with equal diagonal and antisymmetric off-diagonal components:  
\begin{eqnarray}\label{eq:fvertex_mass}
    M_{\parallel}(\omega)&=&\Phi_0 e n\tau \frac{\omega_0 \tau}{(1-i\omega\tau)^2+(\omega_0 \tau)^2},  \nonumber \\
    M_{\perp}(\omega)&=& {\text{sgn} (q B)}\Phi_0 e n\tau \frac{1-i\omega\tau}{(1-i\omega\tau)^2+(\omega_0 \tau)^2},         
\end{eqnarray}
which relate the vortex momentum $\vect{p} = \mat{M}\,\vect{v}$ in the plane perpendicular to the magnetic field direction, $\vect{B}= B \vect{z}$, to the vortex velocity \vect{v}~\cite{Kopnin1995,Kopnin1998}.
Here $\tau$ is the lifetime of the inter CdGM excitation $\omega_{0}$, $n$ carrier density in the vortex core, and $\Phi_0 = h/(2e)$ the magnetic flux quantum. 
The factor ${\text{sgn} (q B)}$ defines the vortex helicity~\cite{sgn}.
Both diagonal and off-diagonal masses are complex, reflecting the retardation between vortex motion and the time-dependent response of quasiparticles in the core. 

This momentum–velocity relation is diagonal in helical basis. 
In a multiband superconductor, the vortex momentum decomposes into independent helical contributions from each band,
$p_\pm = v_\pm \sum_k M_\pm^k$,
where $M_\pm^k$ carries both the band index $k$ and optical chirality, 
\begin{eqnarray} \label{eq:helical_mass}
    M_{\pm}^k(\omega)& = & M_{\parallel}^k \mp i\,M_{\perp}^k\\
    &=& \frac{\Phi_0 e  n_k\tau_k \, e^{i \arctan \frac{\mp {\text{sgn}(q_k B)}}{\tau_k (\omega_0^k \pm{\text{sgn}(q_k B)} \omega)}}}{\sqrt{1 + (\tau_k)^2 (\omega \pm {\text{sgn}(q_k B)}\, \omega_0^k)^2}}. \nonumber         
\end{eqnarray}
Here $m_k^{*}$ is the effective mass of band $k$, $q_k$ the carrier charge and $n_k$ carrier density in this band. 
This expression clearly shows that for a given field direction each band contributes only to one helicity.

In the static limit, $\omega \rightarrow 0$, the two helical masses are equal
\begin{equation}\label{eq:helical_mass_zero_f}
    |M_{\pm}^k(0)| =  \frac{\Phi_0 e  n_k\tau_k }{\sqrt{1 + (\tau_k \omega_0^k)^2 }}.          
\end{equation} 
In the superclean limit, $ \omega^k_{0} \tau_k \gg 1$, the static value 
$|M^k_{\pm}(0)|  \approx 2 \pi (\xi_0^k)^2 m_k^{*}  n_k $
is the linear density of the superfluid mass inside the vortex core and  $\xi_0^k$ superconducting coherence length in band $k$. 
Here we used $\omega^{k}_0  \approx B^{k}_{c2} e/ m^*_{k}$ and 
$B^{k}_{c2} \approx \Phi_0[2 \pi (\xi_{0}^{k})^2]^{-1} $.

\begin{figure}[t]
       \includegraphics[width=.9\columnwidth]{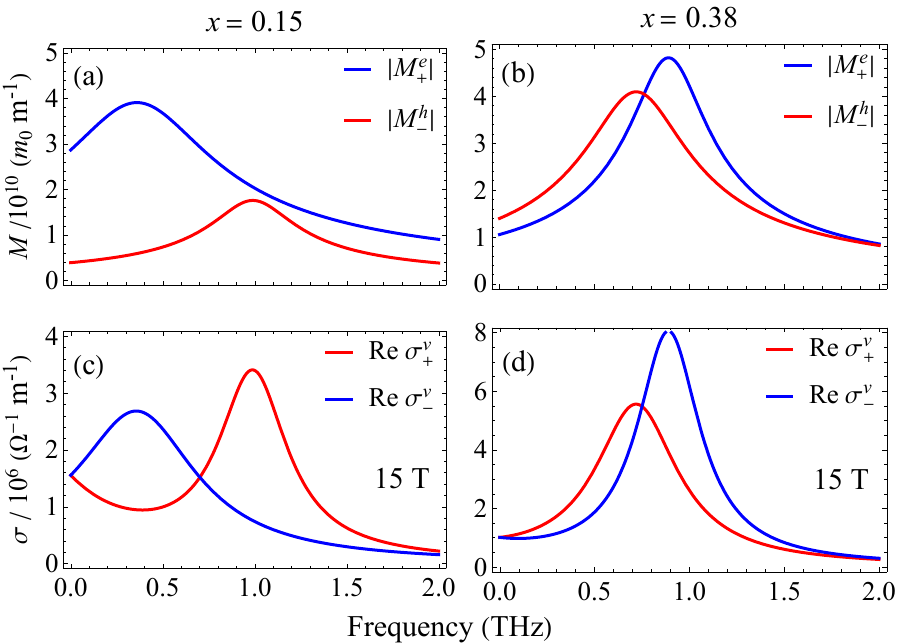}
    \caption{    \label{fig:sigma_mass}
The frequency-dependent vortex response in a two-component superconductor.  
(a, b) -- electron and hole helical vortex mass amplitudes $|M^{k}_{\pm}(\omega)|$, Eq.\,(\ref{eq:helical_mass}); and 
(c, d) -- vortex conductivity $\Re{\,\sigma_{\pm}^{v}}$ in 15\,T, Eq.\,(\ref{eqn:sigma_vortex}); 
Parameters from Table\,\ref{tab:fit_values}, (a, c) for $x=0.15$ and (b, d) for $x=0.38$.  
    }
    \vspace{-5mm}
\end{figure}

Fig.\,\ref{fig:sigma_mass} visualizes how the helicity-selective vortex inertia directly translates into helicity-resolved optical conductivity.
In panels (a) and (b), we plot the amplitudes $|M^{e}_{-}|$ and $|M^{h}_{+}|$ for electron-like and hole-like carriers, respectively. 
These amplitudes are calculated using the parameters resulting from the fits to the Faraday rotation data shown in Fig.\,\ref{fig:thetaF_B_T_dep}, as discussed below, and listed in Table\,\ref{tab:fit_values}.
For electron-like carriers, the resonance at the $\omega^{e}_{0}$ frequency appears only in the left-handed vortex mass per unit length $|M^{e}_{+}(\omega)|$. 
In contrast, hole-like carriers only contribute to the resonance in the right-handed $|M^{h}_{-}(\omega)|$ at the $\omega^{h}_{0}$ frequency. 
For all cases $\omega^{k}_0 \tau_k \gtrsim 1$, putting these superconductors in moderately clean limit with well defined resonances.

\emph{Optical conductivity.} 
The Faraday response is governed by the combination of superfluid and vortex-core conductivities. 
Because an applied $ac$ electric field $E_\pm(\omega)$ at frequency $\omega$ drives both the superconducting current $J_\pm$ and vortex motion in series, the total inverse optical conductivity of a superconductor in the vortex state is given by sum~\cite{Kopnin1998,Kopnin2001,Kopnin2002,Tesar2021,Tesar2024}:
    \begin{equation}\label{eq:inverse_sigma}
            \frac{E_\pm}{J_\pm} = \frac{1}{\sigma_{\pm}} = \frac{1}{\sigma^{sc}} + \frac{1}{\sigma_{\pm}^{v}},  
    \end{equation}
where the  first term contains inverse superfluid conductivity response outside the vortex cores,
$\sigma^{sc}(\omega)=i\epsilon_0\,\omega_{ps}^2/\omega + \sigma_0$, 
with the superfluid plasma frequency, $\omega_{ps}^2=\sum_k(\omega_{ps}^k)^2$, and Drude conductivity of normal carriers present in films, $\sigma_0$~\cite{sigma0}. 

The inverse of the second term in Eq.\,(\ref{eq:inverse_sigma}) is the vortex conductivity
    \begin{align}
    \label{eqn:sigma_vortex}
     \sigma_{\pm}^{v}(\omega,B) & = \frac{1}{|B|\,\Phi_0} \times \nonumber \\
     & 
      \left[ \sum_{k} [(1 - i\omega \tau_k)\frac{M^k_{\pm}}{\tau_k} \pm i \Phi_0 e n_k ]+  i\frac{\kappa}{\omega}\right] 
    \end{align}
which scales with the vortex density and governs the dissipative response in the Abrikosov phase with pinned vortices. 
The sum in Eq.\,(\ref{eqn:sigma_vortex}) runs over all $k$ bands, and $\kappa$ is distribution average of single-vortex Labusch pinning (spring) constant~\cite{Labusch_1968,Beek1993,Moshchalkov_2003, kappa}. 
 
The $\Re{\sigma_{\pm}^{v}(\omega)}$ is shown in Fig.\,\ref{fig:sigma_mass}(c) and (d) for the parameters of Table\,\ref{tab:fit_values}.
The resonant structure arises exclusively from the frequency-dependent vortex mass associated with CdGM transitions~\cite{res_struc}. 
As expected, $\Re{\sigma_{+}^{v}(\omega)}$ is dominated by the resonance of the hole carriers $\omega_0^{(h)}$, while $\Re{\sigma_{-}^{v}(\omega)}$ is dominated by the resonance of the electron carriers $\omega_0^{(e)}$. 

\emph{Model for Faraday rotation induced by a superconducting film on a substrate.} 
Next we simultaneously fit the entire set of field-dependent Faraday rotation data shown in Fig.\,\ref{fig:thetaF_B_T_dep}. 

For small rotation angles $\theta_F$ can be calculated as \cite{Levallois2015}
\begin{equation}
\label{eq:fit_function}
\theta_F(\omega,B) = \frac{1}{2} \mathrm{Arg}\left(\frac{t_-}{t_+}\right),
\end{equation}
where $t_\pm$ are the transmission coefficients for two circular polarizations which can be expressed by the Glover-Tinkham Eq.\,(\ref{eq:tpm}) via optical conductivity of the superconducting film $\sigma_{\pm}$, Eq.\,(\ref{eq:inverse_sigma}), which takes into account the Fabry-Perot resonances in the substrate~\cite{TSS}.

\begin{table}[b]
\vspace{-5mm}
\caption{
The parameters of fit to Faraday rotation data, uncertainty of the last significant digit in the parentheses, and estimated parameters for \fts{x} films. 
$|M(0)_\pm|$ is electron and hole part of the dc values of the vortex mass per unit length modulus, $m_0$ is the bare electron mass.
}
\label{tab:fit_values}
\begin{ruledtabular}
\begin{tabular}{l|cc|cc}
& \multicolumn{2}{c|}{$x=0.15$ }         & \multicolumn{2}{c}{$x=0.38$ }  \\
 & electrons & holes  & electrons & holes \\ \hline
$T_c$ (K)                              &\multicolumn{2}{c|}{$12.8$}                   &\multicolumn{2}{c}{$12.9$}\\
$n / 10^{27}$\,(m$^{-3}$)              &  0.26(6) & $0.07(1)$        & 0.17(4) & $0.19(5)$\\
$\tau$\,(ps)                               & $0.41(1)$   & $0.71(1)$    & $0.80 (2)$  & $0.61(3)$ \\
$\omega_0/(2 \pi)$\,(THz)  & $0.36(2) $    & $0.99(3)$  & $0.89(9)$  & $0.72(7) $ \\
$\omega_{ps}/(2\pi)$\,(THz)   & \multicolumn{2}{c|}{$34.1(4)$}        & \multicolumn{2}{c}{$66.2(9)$} \\
$\sigma_0/10^5$\,($\Omega^{-1} \text{m}^{-1}$)   & \multicolumn{2}{c|}{$2.9(1)$}        & \multicolumn{2}{c}{$2.9(4)$} \\
$\kappa$/$10^4$ (N\,m$^{-2})$                       &\multicolumn{2}{c|}{ $4.6(2) $}         & \multicolumn{2}{c}{$2.0(1)$} \\ 
$\omega_0 \tau$                                 & 0.93  & 4.42             & 4.47  & 2.76 \\ 
$d B_{c2}/d T|_{T \to T_c}$\,(T/K) & \multicolumn{2}{c|}{$6.38$}        & \multicolumn{2}{c}{$7.36$} \\
$m^*/m_0$                       & 3.4  & 3.3             & 3.4  & 3.3 \\
$B_{c2}$\,(T)                      & 44  & 116             & 108  & 85 \\
$\xi_{0}$\,(nm)                      & 2.7  & 1.7             & 1.7  & 2.0 \\
$|M_\pm(0)|/m_0/10^{10}\,(\mathrm{m}^{-1})$   & $2.9$   & $0.39$          & $1.1$   & $1.4$ \\ 
\end{tabular}
\end{ruledtabular}
\end{table}
The best fit parameters to the Faraday rotation data along with the film properties  estimated from the fit results are collected in Table\,\ref{tab:fit_values}. 

The CdGM resonance frequencies $\omega^{k}_0$ are comparable for electrons and holes in the film $x=0.38$. 
In contrast, for $x=0.15$ film $\omega^{e}_0$ is about 3 times lower than $\omega^{h}_0$. 

The superconducting plasma frequency $\omega_{ps}$ 
incorporates the contributions from all the bands~\cite{sigma0},  
and is consistent with the  optical investigations on  single crystals~\cite{Homes2015} and films~\cite{RomeroIII2026}.
The optics data also provides an estimate for the normal state scattering time $\tau_{tr} \approx 0.15$\,ps\,\cite{Homes2015,RomeroIII2026,Stensberg_persistent_2025}.  

The derived parameters place both films in the moderately clean limit, mean free path $l\gtrsim\xi_0$~\cite{Kopnin1996}, and $\omega_0\tau \gtrsim 1$ for all considered bands. 
The parameters show a systematic redistribution of superfluid weight between the electron and hole bands with isovalent substitution $x$. 

The sum of carrier densities in the vortex core, $n_e + n_h$, changes only weakly with isovalent substitution $x$. 
However,  the ``chemical pressure" caused by Te/Se substitution  modifies the relative sizes of the electron vs hole pockets: 
while for the $x=0.38$ superconductor the density of hole vs. electrons is about even, and the electrons have slightly longer lifetime. 
For the $x=0.15$ film the electrons dominate the carriers density but have shorter lifetime than holes. 

All obtained parameters are in general agreement with independent measurements: 
the Fermi pocket sizes and effective masses estimated by angle resolved photoemission spectroscopy (ARPES) studies~\cite{Tamai2010,ZXShen_2015,Rinott2017,Coldea_2023}; 
the vortex pinning frequency 
$(2 \pi)^{-1}\sqrt{\kappa/(|M_+^e| (\omega) + |M_-^h| (\omega))} = 0.12 \div 0.17$\,THz  
at $\omega = 0.3$\,THz (Fig.\,\ref{fig:sigma_mass}, Table\,\ref{tab:fit_values}) is similar to 0.15\,THz obtained in \cite{Nakamura2024,mass_nakamura};
the vortex pinning constants by prior microwave and THz second harmonic generation studies~\cite{Okada2015,Pompeo2020}.

The transport measurement of the upper critical field slope near $T_c$ $d B_{c2}/d T|_{T \to T_c}$ allows us to estimate the effective masses $m^*_{k}$ for electron and hole pockets of the most mobile bands using the Prozorov-Kogan formalism~\cite{Prozorov_2024}. 
These masses, approximately 3.3~$m_{0}$, appear to be consistent with the masses obtained from ARPES data for the most mobile bands~\cite{Rinott2017,Tamai2010,ZXShen_2015,Coldea_2023} or by dynamical mean-field theory calculations~\cite{Yin2011}. 

Knowing the CdGM resonance frequencies and effective masses, we can estimate the upper critical field and the coherence length for each band 
$B^{k}_{c2} \approx \omega^{k}_0\, m^*_{k}/e$ and 
$\xi_0^{k} \approx \sqrt{\Phi_0/2 \pi B^{k}_{c2}}$. 
We find that while for $x=0.38$ film both condensates have a similar coherence length, for the $x=0.15$ superconductor the electron condensate has a longer coherence length than the hole condensate. 

We note that all analyses were performed within the framework of noninteracting condensates. 
Intercondensate coupling, neglected here, may hybridize CdGM states with different band indexes~\cite{Berthod2018}, the effects of which could be accessed in future studies.

\emph{Conclusions.} 
We have used terahertz Faraday magneto-optical spectroscopy to resolve helical, quantized CdGM states in nodeless multiband FeTe$_{1-x}$Se$_x$ superconducting films.
The polarization-selective response enables band-resolved access to vortex-core dynamics, revealing distinct electron- and hole-band contributions to the superfluid density, vortex mass, and coherence length. 
We track the evolution of multiband superconductivity as a function of the isovalent substitution $x$. 

Our results demonstrate that terahertz magneto-optics provides a dynamical probe of vortex-core quasiparticles and establish direct evidence for multiband helical CdGM physics in iron-based superconductors. 
The developed magneto-optical approach opens a route to chirality-resolved spectroscopy of vortex matter in complex superconductors and is applicable to any type-II superconductor with sufficiently short coherence length, including iron-based, heavy-fermion, and engineered superconducting heterostructures.

\emph{Acknowledgments.} 
We are grateful for discussions with I.\,Timoshuk and E.\,Babaev, 
and we thank M.\,M{\o}ller for technical help.  
The work at NICPB was supported by the European Research Council (ERC) under the European Union Horizon 2020 research and innovation program Grant Agreement No. 885413. 
The thin film growth at Rutgers was supported by the National Science Foundation Grant No. DMR-2451900. 
The spectroscopic work at Rutgers was supported by the National Science Foundation Grant No. DMR-2105001.


\begin{thebibliography}{76}
\bibitem [{max()}]{max_rot}%
  \BibitemOpen
  \href@noop {} {}\bibinfo {note} {Note that due to screening effects the
  maximum in the Faraday rotation spectrum is at a significantly lower
  frequency than the bare CdGM resonances $\omega_0^{k}$,
  Fig.\,\ref{fig:sigma_mass}}\BibitemShut {NoStop}%
\bibitem [{Ver()}]{Verdet_par}%
  \BibitemOpen
  \href@noop {} {}\bibinfo {note} {Because of contribution of thin film
  interfaces to the rotation the ``effective Verdet parameter'' depends on the
  material thickness.}\BibitemShut {Stop}%
\bibitem [{sgn()}]{sgn}%
  \BibitemOpen
  \href@noop {} {}\bibinfo {note} {The ${\text{sgn} (q)}$ is $+1$ for holes and
  $-1$ for electrons and $\text{sgn}(B)$ is the sign of magnetic field
  projected on the $\mat{z}$ axis.}\BibitemShut {Stop}%
\bibitem [{res()}]{res_struc}%
  \BibitemOpen
  \href@noop {} {}\bibinfo {note} {Note that although the superconducting
  condensate conductivity $\sigma_{sc}$ and the pinning term in vortex
  conductivity $\sigma_\pm^v$ have the same non-resonant frequency dependence,
  $i/ \omega$, the latter is scaled by magnetic field, see
  Eq.\,(\ref{eqn:sigma_vortex}).}\BibitemShut {Stop}%
\end{thebibliography}
%



\newpage
\section{End Matter}
\begin{figure}[b]  
    \includegraphics[width=0.85\columnwidth]{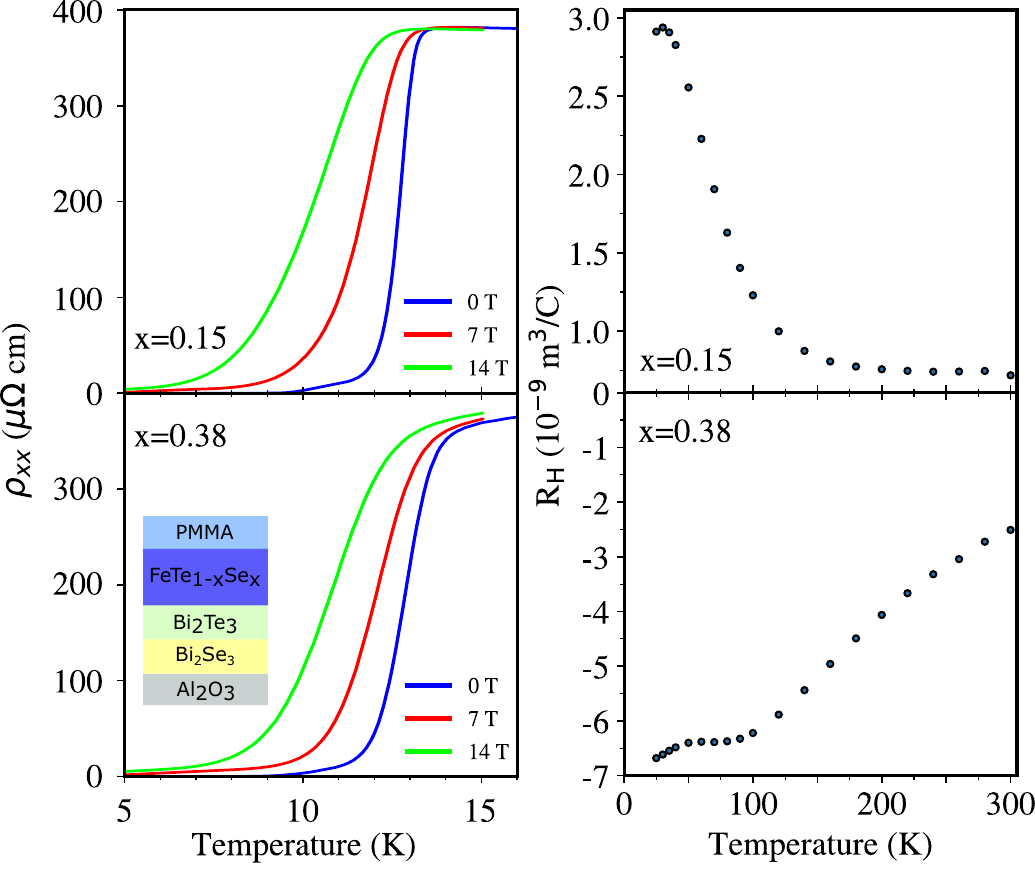}
    \vspace{-2mm}
    \caption{\label{fig:Rxx_RH}
        Temperature dependence of resistivity for samples $x=0.15$ (top) and $x=0.38$ (bottom) at several magnetic fields. 
        The right column shows the temperature dependence of the Hall coefficient $R_H$.
        The sample layer structure can be seen in the inset.
    }
\end{figure}

\vspace{-3mm}
\emph{Sample growth and characterization.} 
Epitaxial \fts{x} films of thickness 40\,nm were grown at nominal doping levels of $x = 0.15$ and $x = 0.38$ by molecular beam epitaxy (MBE)~\cite{SOh_2021,SOh_2025}. 
The heterostructure consisted of a 3\,nm Bi$_2$Se$_3$ buffer layer deposited on a 
 c-cut $10 \times 10$\,mm$^{2}$ 
Al$_2$O$_3$ substrate, followed by a 7\,nm Bi$_2$Te$_3$ layer, the 40\,nm \fts{x} film, and a 300\,nm PMMA capping layer for protection, Fig.\,\ref{fig:Rxx_RH} inset. 
The buffer and capping layers do not contribute measurably to the Faraday response in the superconducting state.

Electrical transport measurements were performed shortly after completion of the THz spectroscopy. 
Fig.\,\ref{fig:Rxx_RH} shows the magnetic-field dependence of the resistivity and the temperature dependence of the Hall coefficient   obtained from $\rho_{xy} = R_HB$. 
The positive $R_H$ for $x=0.15$ indicates hole-dominated conduction, while the negative $R_H$ for $x=0.38$ indicates electron-dominated conduction.

The superconducting transition temperature $T_c$, Table\,\ref{tab:fit_values}, was determined from the peak in the derivative of $\rho_{xx}(T)$. 
The slope of the upper critical field $d B_{c2}/d T|_{T \to T_c}$, Table\,\ref{tab:fit_values}, for field perpendicular to the sample surface was derived from temperature-dependent $\rho_{xx}$ measurements at $B = 0$, 7, and 14\,T. 
The measured slope is consistent with the earlier report~\cite{Sawada2016}.

\emph{Glover-Tinkham transmission expression for film on a substrate.} 
The transmission coefficients in helical coordinates for a superconducting film on a substrate are given by the Glover-Tinkham expression~
\cite{Tinkham_1956,Tinkham1956GapMeasurement,Glover-Tinkham_1957,Heinz_2018,Tanner2019,Ubrig2013}: 
\begin{equation}\label{eq:tpm}
\mathsmaller{
     t_\pm = 
     \frac{ 4 p_2 N_2}{(N_2 + 1)^2 - (N_2 - 1)^2 p_2^2 + Z_0 d_f \sigma_\mp[N_2 + 1 + (N_2-1) p_2^2]}.
}
\end{equation}
Here $N_2=3.14$ is the refraction index of the Al$_2$O$_3$ substrate, $p_2=e^{i N_2 d_s \omega /c}$ is a single passage phase difference for light propagation in the substrate, $d_s$ (0.500\,mm and 0.494\,mm for $x=0.15$ and $x=0.38$  samples, respectively) and $d_f$ are substrate and film thicknesses, $\sigma_\mp(\omega)$ is film optical conductivity, Eq.\,(\ref{eq:inverse_sigma}) and (\ref{eqn:sigma_vortex}), and $Z_0\approx 377\,\Omega$ is the vacuum impedance.

The transmission  through a parallel slab of \fts{x} film of thickness $d_f$ in vacuum is
\begin{equation}\label{eq:slab}
t_\pm=\frac{N_\pm p_\pm}{ (N_\pm +1)^2 - (N_\pm -1)^2 p_\pm^2},
\end{equation}
where $p_\pm = e^{i N_\pm d_f \omega /c}$, $N_\pm$ is the film index of refraction, Eq.\,(\ref{eq:N_index}), $\sigma_\pm$ is the conductivity of the film, Eq.\,(\ref{eq:inverse_sigma}), and  $\epsilon_\infty=4$~\cite{Homes2011}.

\end{document}